# Optical and RF Metrology for 5G

(*Invited*)

*David A. Humphreys,
Irshaad Fatadin*
National Physical Laboratory
Teddington, UK.
david.humphreys@npl.co.uk

*Mark Bieler,
Paul Struszewski*
Physikalisch-Technische
Bundesanstalt
Braunschweig, Germany

*Martin Hudlička*
Czech Metrology Institute
Brno, Czech Republic

*Abstract*— Specification standards will soon be available for 5G mobile RF communications. What optical and electrical metrology is needed or available to support the development of the supporting optical communication systems? Device measurement, digital oscilloscope impairments and improving system resolution are discussed.

## I. Introduction

In Europe the research effort to develop the next generation of mobile system, 5G, is strongly linked to the Horizon 2020 European research programme [1]. 3GPP has defined use-cases [2] based on the worldwide research, leading to the anticipated release of the first set of 5G standards in the second half of 2017 [3]. 5G is primarily a mobile radio access protocol and the primary differences from the current (4G) LTE system are the use of mm-wave frequencies, greater spatial diversity, x 1000 capacity increase and low (1 ms target) latency, so how does optical communications fit in and how does metrology help?

Much of the existing 5G research is associated with the RF aspects such as massive MIMO, Over-The-Air testing and mm-wave propagation. The envisioned increased data capacity mean that it will fall to optical communication technologies to provide the heavy lifting required for the fronthaul. It is likely that an Ethernet-based flexible fronthaul will provide a scalable common platform but the challenges for latency and latency variation remain serious [4]. Given that these systems would be deployed at the edge of the network rather than as core transport, low-cost, high-speed optical communication links and multiplexing technology are essential.

Measurement is important to facilitate trade as well as assisting in research and development. In general, during the initial research phases there is a lack of dedicated test equipment and much of the work is done through simulation or using general test equipment.

Within Europe, Euramet has funded a series of collaborative projects with national measurement Institutes, Industry and Universities with the aim of improving the traceability and uncertainty propagation for dynamic waveform generation and measurement (IND16 "Ultrafast"), RF and optical communications research (IND51 "MORSE"), Metrology for optical component and 5G communications (see funding).

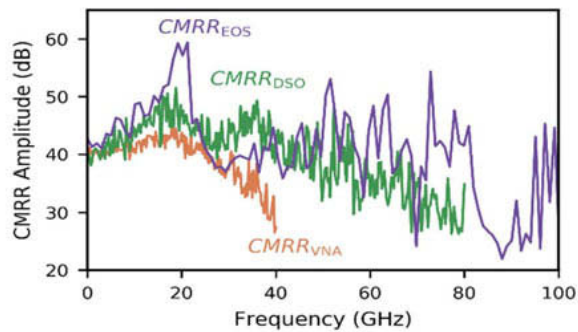

Fig. 1. Comparison of the CMRR using the EOS, DSO and VSA (see [8]).

## II. High-speed electrical and optoelectronic measurement

The front-haul optical communications requirement for 5G will make use of the current communication sub-systems.

Traceability for dynamic measurements is through electro-optic sampling. Systems are available at and have been cross-validated by NPL, PTB and NIST. Intercomparison is important as measurement systems will contain some imperfections that can be quantified. This is also true for commercial test equipment and so we have also developed algorithmic corrections that can be applied to minimize these errors. One of the most important recent developments is the use of the optoelectronic pulse generation to quantify errors in the microwave mismatch of the device under test by measuring at several points on the transmission line to separate the forward and reverse travelling-waves [5]. Previously, the impedance mismatch correction was made using a Vector Network Analyzer, which has an abrupt frequency cut-off [6]. The new approach, overcomes this difficulty as the pulse will contain higher-frequency components, giving a time-domain result that is not masked by truncation errors.

### A. Photodiode response and Common-Mode Rejection Ratio

Photodiodes can be directly calibrated using electro-optic sampling but this is a slow process so it is more common to use calibrated digital sampling oscilloscopes (DSO) or digital real-time oscilloscopes (DRTO). The fidelity of the DSO timebase is poor in comparison with the DRTO. Either a precision

EMPIR projects 14IND13 PhotInd and 14IND10 MET5G. The EMRP was supported by the EMRP participating countries within EURAMET and the European Union. The EMPIR program has been co-financed by the Participating States and from the European Union's Horizon 2020 research and innovation timebase or two instrument channels can be used with correction algorithms [7] to overcome this problem.

Common-Mode Rejection-Ratio (CMRR) is an important parameter for dual photodiodes that are a key component of the coherent detection optical hybrid [8]. The standard definition for CMRR has been modified by adding scaling ($\alpha$) and time-shift ($\tau$) parameters which must be optimized:

$$CMRR = 20 \log \left( \left| \frac{\alpha \exp(j\omega\tau)V_p(f) - V_n(f)}{\alpha \exp(j\omega\tau)V_p(f) + V_n(f)} \right| \right) \quad (1)$$

Where $V_p$ and $V_n$ are the signals from the positive and negative photodiodes respectively. Our results showed that for an EOS or Vector Network Analyzer (VNA) measurement this approach gave a satisfactory result but when using a DSO as the electrical measuring system it was necessary to illuminate both photodiodes simultaneously and balance the optical powers to minimize the residual response. It was also noted that balancing the photocurrents did not correspond to the optimum CMRR value. The inference is that if an active alignment process is used then it may be better if an RF modulated optical signal is used.

The impulse response measurements were made using a 1500 nm modelocked laser system whereas the frequency-domain measurements were made using a VNA and 20 GHz LiNbO$_3$ intensity modulator. The results, summarized in Fig. 1 are similar but there is clear degradation of the VNA results due to the frequency-response roll-off. The important points to note are that the CMRR is a ratiometric measurement the absolute response of the measurement system only affects the bandwidth and noise performance. The high CMRR >40 dB over the full bandwidth suggest that the absorption profile drift components of the photodiodes' response are well matched.

Although in this case the VNA results were degraded a system of this type can be constructed using commercial components and coupled with a probe-station for active device alignment. Such a system is robust and would be ideal for a production environment.

## III. MODULATED OPTICAL AND ELECTRICAL WAVEFORM MEASUREMENT

Measurement and assessment of the modulator and receiver systems during development rely on state-of-the-art DRTO and Arbitrary Waveform Generators (AWG). The move from simple modulation formats to higher-order constellations brings a greater requirement for linearity of the driver and receiver electronics. Error-Vector Magnitude (EVM) is a parameter that is often used to monitor the waveform error and this can be tied to the predicted bit-rate [9] (Fig. 2). EVM and other digital metrics are currently being studied by IEEE P1765.

Typically the AWG and DSO may have different clock rates. As the sampling is strict, the representation of the waveform may be mathematically sufficient but sparse. For example an AWG generating symbols at 28 GBaud and sampled at 100 GSa/s will be represented by 3.57 points per symbol. The effective sampling-rate can be improved using the highest common factor for the two sample rates and a suitable choice of PRBS sequence length (Fig. 3).

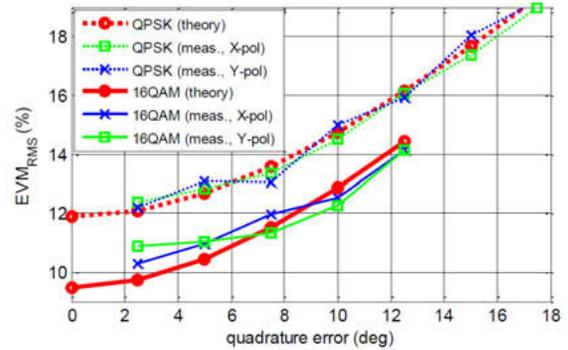

Fig. 2. Agreement between the theoretical EVM (6) and measured values, determined by error counting and calculated from EVM (see [9]).

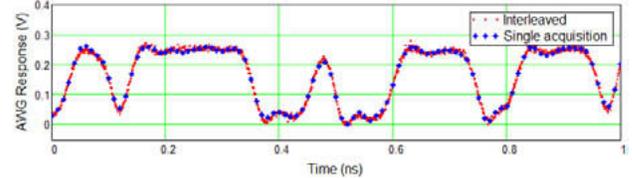

Fig. 3. Single and interleaved acquisitions using the prime-number relationship between the AWG and DRTO sampling rates

ACKNOWLEDGMENT

David Humphreys thanks Nathan Gomes for providing material from his 5G research and Marco Peccianti and Hualong Bao for the use of their fs laser system.